\documentclass[12pt,dvips]{article}

\usepackage{rotating}
\usepackage{hhline}
\usepackage{array}
\usepackage{epsfig}

\begin{document}

\renewcommand{\thefootnote}{\fnsymbol{footnote}}
\tolerance=100000

\newcommand{\imag}{\Im {\rm m}}
\newcommand{\real}{\Re {\rm e}}
\newcommand{\s}{\\ \vspace*{-3.5mm}}

\def\tablename{\bf Table}%
\def\figurename{\bf Figure}%


\begin{flushright}
KIAS-P03064 \\[-0.1cm]
hep-ph/0308060\\
August 2003
\end{flushright}

\vspace{1.2cm}

\begin{center}
  {\Large \bf Neutralino Pair Production and 3--Body Decays at
              \boldmath{$e^+e^-$} Linear Colliders as Probes of CP Violation
              in the Neutralino System}\\[1.5cm]
  {\large  S.Y. Choi}\\[0.3cm]
  {\it Department of Physics, Chonbuk National University,
       Chonju 561-756, Korea}
\end{center}

\renewcommand{\thefootnote}{\arabic{footnote}}
\vspace{3.cm}

\begin{abstract}
\noindent In the CP--invariant supersymmetric theories, the steep
S--wave (slow P--wave) rise of the cross section for any
non--diagonal neutralino pair production in $e^+e^-$ annihilation,
$e^+e^-\rightarrow \tilde{\chi}^0_i\tilde{\chi}^0_j$ ($i\neq j$),
near threshold is accompanied by the slow P--wave (steep S--wave)
decrease of the fermion invariant mass distribution of the 3--body
neutralino decay, $\tilde{\chi}^0_i\rightarrow\tilde{\chi}^0_j\,
f\bar{f}$ ($f=l$ or $q$), near the end point. These selection
rules, unique to the neutralino system due to its Majorana nature,
guarantee that the observation of simultaneous sharp S--wave
excitations of the production cross section near threshold and the
lepton or quark invariant mass distribution near the end point is
a qualitative, unambiguous evidence for CP violation in the
neutralino system.
\end{abstract}
%


\newpage

\section{Introduction}

Most supersymmetric extensions of the Standard Model (SM) based on
some soft supersymmetry (SUSY) breaking mechanism contain several
CP phases, whose large values tend to render lepton and quark
electric dipole moments (EDM) too large to satisfy stringent
experimental constraints \cite{susycp}. Such CP crises are generic
in supersymmetric theories, but may be resolved by pushing the
masses of some sparticles, especially the first and second
generation sfermions, above a few TeV, by arranging for internal
cancellations, or by simply setting phases to be extremely small
\cite{overcome}. On the other hand, new sources of CP violation
beyond the SM are required to explain the non--zero baryon
asymmetry in the universe in the standard Big Bang framework
\cite{baryon}. Therefore, it is  crucial to look for new
signatures for CP violation in such SUSY scenarios with some large
phases, as long as they are consistent with the stringent EDM and
other low--energy constraints. In this light, detailed analyses of
the neutralino sector at future $e^+e^-$ linear collider
experiments \cite{futurecoll} can prove particularly fruitful
\cite{petcov,ckmz,Gudi,choi,11a}, because in most supersymmetric
theories neutralinos belong to the class of the lighter
supersymmetric particles \cite{snowmass} and the neutralino system
contains two non--trivial CP violating phases.\s

There are many different ways for probing CP violation in the
neutralino system. The imaginary parts of the complex parameters
in the neutralino mass matrix could most directly and
unambiguously be determined by measuring suitable {\cal CP}
violating observables by exploiting initial beam polarization and
angular correlations between neutralino production and decay at
future high--energy colliders \cite{ckmz,Gudi,choi,11a}. But,
their experimental measurements will be quite difficult. The
presence of the CP violating phases can also be identified through
by their impact on CP--even quantities such as neutralino masses,
branching ratios and so on. However, since these quantities are
already non--zero in the CP conserving case, the detection of the
presence of non--trivial CP phases will require a careful
quantitative analysis of a number of physical observables,
especially for small CP--odd phases giving rise to very small
deviations from the CP--conserving values \cite{susycp}. On the
other hand, the rise of excitation curves near threshold for
non--diagonal neutralino pair production in $e^+e^-$ collision is
altered qualitatively in CP--noninvariant theories
\cite{petcov,ckmz}, by allowing the steep S--wave increase of all
pairs simultaneously. Thus, as demonstrated in Ref.~\cite{ckmz},
precise measurements of the threshold behavior of the
non--diagonal neutralino pair production processes may give clear
indications of non--zero CP violating phases in the neutralino
sector, if at least three different neutralino states are
accessible kinematically.\s

In the present note we provide a new powerful method for probing
CP violation in the neutralino system, which is based on a
combined analysis of the threshold excitations of neutralino pair
production in $e^+e^-$ annihilation and the fermion invariant mass
distribution near the end point of the 3--body neutralino
fermionic decays:
\begin{eqnarray*}
e^+e^-\,\rightarrow\, \tilde{\chi}^0_i\tilde{\chi}^0_j\quad (i\neq
j) \quad {\rm and}\quad \tilde{\chi}^0_i\,\rightarrow\,
\tilde{\chi}^0_j\, f\bar{f}\quad (f=l, q)\,.
\end{eqnarray*}
[The 3--body decay process includes clean $\mu^+\mu^-$ and
$e^+e^-$ decay channels with little background, which allow a
clear reconstruction of the kinematical configuration with good
precision.] This method relies on selection rules, unique to the
neutralino system due to its Majorana nature in CP--invariant
theories, and it can work effectively if the branching ratios of
the 3--body neutralino fermionic decays are not suppressed. [Once
two--body decays of the neutralino $\tilde{\chi}^0_i$  into $Z$,
Higgs bosons or sfermions are open, the new method is
ineffective.]\s

Before demonstrating the new  method for probing CP violation in
the neutralino system in detail, we describe briefly the mixing
for the neutral gauginos and higgsinos in CP--noninvariant
theories with non--vanishing phases in Sec.~\ref{sec:mixing}. In
Sec.~\ref{sec:threshold} we introduce the selection rules for the
production of neutralino pairs and the neutralino to neutralino
transition via a (virtual) vector boson or sfermion exchange.
Then, we prove that in any CP--invariant SUSY theory, if the
production cross section for any non--diagonal neutralino pair in
$e^+e^-$ annihilation increases steeply in S--waves (slowly in
P--waves) near threshold, the lepton or quark invariant mass
distribution of the decay
$\tilde{\chi}^0_i\rightarrow\tilde{\chi}^0_j\, f\bar{f}$ ($f=l$ or
$q$) decreases slowly in P--waves (steeply in S--waves) near the
end point. Thus, the observation of simultaneous sharp S--wave
excitations of both the production of any non--diagonal neutralino
pair $\tilde{\chi}^0_i\tilde{\chi}^0_j$ near threshold and the
fermion invariant mass distribution of the decay
$\tilde{\chi}^0_i\rightarrow\tilde{\chi}^0_j\, f\bar{f}$ near the
end point will be a qualitative, unambiguous evidence for CP
violation in the neutralino system. A quantitative demonstration
of the method based on a specific set of the relevant
supersymmetry parameters is given in the last part of
Sec.~\ref{sec:threshold}. Finally, conclusions are drawn in
Sec.~\ref{sec:conclusion}.

\section{Neutralino Mixing}
\label{sec:mixing}

In the minimal supersymmetric extension of the Standard Model
(MSSM), the mass matrix of the spin-1/2 partners of the neutral
gauge bosons, $\tilde{B}$ and $\tilde{W}^3$, and of the neutral
Higgs bosons, $\tilde H_1^0$ and $\tilde H_2^0$, takes the form
\begin{eqnarray}
{\cal M}=\left(\begin{array}{cccc}
  M_1  &  0   &  -m_Z c_\beta s_W  &  m_Z s_\beta s_W \\[2mm]
   0   & M_2  &   m_Z c_\beta c_W  & -m_Z s_\beta c_W\\[2mm]
-m_Z c_\beta s_W & m_Z c_\beta c_W &  0   & -\mu  \\[2mm]
 m_Z s_\beta s_W &-m_Z s_\beta c_W & -\mu &  0
               \end{array}\right)\,,
\label{eq:massmatrix}
\end{eqnarray}
in the $\{\tilde{B},\tilde{W}^3,\tilde{H}^0_1,\tilde{H}^0_2\}$
basis. Here $M_1$ and $M_2$ are the fundamental supersymmetry
breaking U(1) and SU(2) gaugino mass parameters, and $\mu$ is the
higgsino mass parameter. As a result of electroweak symmetry
breaking by the vacuum expectation values of the two neutral Higgs
fields $v_1$ and $v_2$ ($s_\beta =\sin\beta$, $c_\beta=\cos\beta$
where $\tan\beta=v_2/v_1$), non--diagonal terms proportional to
the $Z$--boson mass $m_Z$ appear and  the gauginos and higgsinos
mix to form the four neutralino mass eigenstates
$\tilde{\chi}_i^0$ ($i=1$--$4$). In general the mass parameters
$M_1$, $M_2$ and $\mu$ in the neutralino mass matrix
(\ref{eq:massmatrix}) can be complex. By re--parameterization of
the fields, $M_2$ can be taken real and positive, while the U(1)
mass parameter $M_1$ is assigned the phase $\Phi_1$ and the
higgsino mass parameter $\mu$ the phase $\Phi_\mu$.\s

The neutralino mass eigenvalues $m_i\equiv m_{\tilde{\chi}^0_i}$
$(i=1,2,3,4)$ can  be chosen positive by a suitable definition of
the mixing matrix $N$, rotating the gauge eigenstate basis
$\{\tilde{B},\tilde{W}^3,\tilde{H}^0_1,\tilde{H}^0_2\}$ to the
mass eigenstate basis of the Majorana fields $\tilde{\chi}^0_i$
($i=1$--4). In general the matrix $N$ involves 6 angles and 10
phases, and can be written as \cite{ckmz,six}
\begin{eqnarray}
N = {\sf diag}\left\{{\rm e}^{i\alpha_1},\,
                   {\rm e}^{i\alpha_2},\,
                   {\rm e}^{i\alpha_3},\,
                   {\rm e}^{i\alpha_4}\,\right\} {\sf R}_{34}\, {\sf
                   R}_{24}\,{\sf R}_{14}\,{\sf R}_{23}\,{\sf
                   R}_{13}\, {\sf R}_{12}\,,
\label{eq:Mdef}
\end{eqnarray}
where ${\sf R}_{jk}$ are rotations in the complex [$jk$] plane
characterized by a mixing angle $\theta_{jk}$ and a (Dirac) phase
$\beta_{jk}$. One of (Majorana) phases $\alpha_i$ is nonphysical
and, for example, $\alpha_1$ may be chosen to vanish. None of the
remaining 9 phases can be removed by rotating the fields since
neutralinos are Majorana fermions. The neutralino sector is CP
conserving if $\mu$ and $M_1$ are real, which is equivalent to
$\beta_{ij}=0$ (mod $\pi$) and $\alpha_i=0$ (mod $\pi/2$).
Majorana phases  $\alpha_i=\pm \pi/2$ do not signal CP violation
but merely indicate different intrinsic CP parities of the
neutralino states in CP--invariant theories \cite{R2}.

\section{Neutralino Pair Production and 3--Body Decays}
\label{sec:threshold}

Both the production processes, $e^+e^-\rightarrow
\tilde{\chi}^0_i\tilde{\chi}^0_j$ ($i,j=1$--4), and the 3--body
neutralino decays, $\tilde{\chi}^0_i\rightarrow \tilde{\chi}^0_j\,
f\bar{f}$, are generated by the five mechanisms: $s$--channel $Z$
exchange, and $t$-- and $u$--channel $\tilde{f}_{L,R}$ exchanges
with $\tilde{f}=\tilde{e}$ for the production processes. After
appropriate Fierz transformations of the sfermion exchange
amplitudes and with the fermion masses neglected, the transition
matrix element of the production process $e^+e^-\rightarrow
\tilde{\chi}^0_i\tilde{\chi}^0_j$ and that of the 3--body
fermionic neutralino decays $\tilde{\chi}^0_i\rightarrow
\tilde{\chi}^0_j\, f\bar{f}$ can be written as
\begin{eqnarray}
T(e^+e^-\rightarrow\tilde{\chi}^0_i\tilde{\chi}^0_j) &=&
\sum_{\alpha,\beta\,=L,R} Q_{\alpha\beta}
  \left[\bar{v}(e^+)\gamma_\mu P_\alpha u(e^-)\right]\,
  \left[\bar{u}(\tilde{\chi}^0_i)\gamma^\mu P_\beta\, v(\tilde{\chi}^0_j)\right]
\label{eq:amplitude1}\,, \\
D(\tilde{\chi}^0_i\rightarrow \tilde{\chi}^0_j\, f\bar{f}) &=&
\sum_{\alpha,\beta\,=L,R} Q'_{\alpha\beta}\,
   \left[\bar{u}(f)\gamma^\mu P_\alpha\, v(\bar{f})\right]\,
   \left[\bar{u}(\tilde{\chi}^0_j)\gamma_\mu P_\beta\, u(\tilde{\chi}^0_i)\right]\,,
\label{eq:amplitude2}
\end{eqnarray}
that is to say, as a sum of the products of a $\tilde{\chi}^0$
vector or axial vector current and a fermion vector or axial
vector current, respectively. We refer to Ref.~\cite{ckmz} and
Ref.~\cite{choi} for the expressions of the generalized bilinear
charges $Q_{\alpha\beta}$ and $Q'_{\alpha\beta}$, just mentioning
that the bilinear charges become independent of the kinematical
variables when two neutralinos are at rest. Therefore, in this
static limit, both the production and the decays can be considered
to proceed via a static vector boson exchange.\s

Some general properties of the bilinear charges $Q_{\alpha\beta}$
and $Q'_{\alpha\beta}$ in Eqs.~(\ref{eq:amplitude1}) and
(\ref{eq:amplitude2}) can be derived in CP--invariant theories by
applying CP invariance and the Majorana condition for neutralinos
to the transition matrix elements. In CP--invariant theories, the
production of a neutralino pair through a vector or axial vector
current with positive intrinsic CP parity satisfies the CP
relation \cite{11a,kayser}
\begin{eqnarray}
1=\eta^i\eta^j (-1)^L\,, \label{cpparity1}
\end{eqnarray}
in the non--relativistic limit of two neutralinos, where
$\eta^i=\pm\, i$ is the intrinsic CP parity of $\tilde{\chi}^0_i$
and $L$ is the orbital angular momentum of the neutralino pair.
The selection rule (\ref{cpparity1}) reflects the fact that if two
neutralinos $\tilde{\chi}^0_i$ and $\tilde{\chi}^0_j$ have the
same or opposite CP parity, the current for the neutralino pair
production must be pure axial--vector or pure vector form,
respectively, cf.~\cite{kayser}. Because the axial--vector current
and the vector current involve the combination of $u$ and $v$
spinors for the two Majorana particles, the axial vector
corresponds to the P--wave ($L=1$) and the vector to the S--wave
($L=0$). \s

On the other hand, the neutralino decay,
$\tilde{\chi}^0_i\rightarrow \tilde{\chi}^0_j + V$, where $V$
stands for the final fermion current in Eq.~(\ref{eq:amplitude2}),
satisfies the CP relation
\begin{eqnarray}
\eta^i=\eta^j (-1)^L \quad {\rm or\ \ equivalently} \quad 1 =
-\eta^i\eta^j (-1)^L\,, \label{cpparity2}
\end{eqnarray}
in the non--relativistic limit of two neutralinos, where $L$ is
the orbital angular momentum of the final state of
$\tilde{\chi}^0_j$ and $V$. We emphasize first that the neutralino
to neutralino transition current is pure axial--vector or pure
vector form for the two neutralinos of the same or opposite CP
parity, respectively, as in the production case. However, because
two $u$--spinors are associated with the currents in the
neutralino to neutralino transition, the axial--vector corresponds
to S--wave excitation while the vector corresponds to P--wave
excitation, giving rise to the relative minus sign between
(\ref{cpparity1}) and (\ref{cpparity2}).\s

One immediate consequence of the selection rules (\ref{cpparity1})
and (\ref{cpparity2}) is that, in CP--invariant theories, if the
production of a pair of neutralinos with the same (opposite) CP
parity through a vector or axial vector current is excited slowly
in P-waves (steeply in S--waves) \cite{R2}, then the neutralino to
neutralino transition via such a vector or axial vector current is
excited sharply in S--waves (slowly in P--waves). More explicitly,
the power of the selection rules (\ref{cpparity1}) and
(\ref{cpparity2}) can clearly be seen by inspecting the
expressions for the S--wave excitations of the total cross section
$\sigma{\{ij\}}$ ($i\neq j$) near threshold and of the fermion
invariant mass distribution of the 3--body neutralino decay
$\tilde{\chi}^0_i\rightarrow\tilde{\chi}^0_j\, f\bar{f}$ (with the
fermion masses neglected) near the end point:
\begin{eqnarray}
\sigma {\{ij\}} & \approx &  \frac{4\pi\alpha^2\, m_i
m_j}{(m_i+m_j)^4}\,
          \beta \,\bigg\{\,|\,\imag\, G_R|^2\, + |\,\imag\, G_L|^2 \bigg\}
          + {\cal O}( \beta^3)\label{eq:thres1}\,,
          \\[0.2cm]
\frac{d\Gamma{\{ij\}}}{d z_{ff}} & \approx &
\frac{2\alpha^2}{\pi}\,
          \left(\frac{m_j}{m_i}\right)^{3/2}\, \left(m_i-m_j\right)\,
          \beta^\prime\,\bigg\{\, |\,\real\, G^{\prime}_R|^2\,
         +  |\,\real\, G^{\prime}_L|^2\bigg\} + {\cal O}(\beta^{\prime 3})\,,
\label{eq:thres2}
\end{eqnarray}
where $\beta =\sqrt{1-(m_i+m_j)^2/s}$ and $\beta^\prime=
\sqrt{1-z^2_{ff}}$ with the dimensionless variable
$z_{ff}=m_{ff}/m^{\rm max}_{ff}$, the ratio of the fermion
invariant mass $m_{ff}$ to its maximal value $m^{\rm
max}_{ff}=m_i-m_j$. Here, the coupling dependent parts, each of
which is connected with the chirality of the neutralino current,
are given by
\begin{eqnarray}
&& G^{(\prime)}_R = -\frac{Q_f}{2c_W^2}\, D^{(\prime)}
                    (N_{i3}N^*_{j3}-N_{i4}N^*_{j4})
                  - \frac{Q^2_f}{c_W^2}F^{(\prime)}_{R} N_{i1}N^*_{j1}\,,
                    \nonumber \\
&& G^{(\prime)}_L = \frac{(I^f_3-Q_f\,s^2_W)}{2c_W^2 s_W^2}\,
D^{(\prime)}
                    (N_{i3}N^*_{j3}-N_{i4}N^*_{j4})
                    + \frac{1}{s^2_Wc^2_W} F^{(\prime)}_{L}N'_{i2}N^{'*}_{j2}\,,
\end{eqnarray}
with $N^\prime_{i2} = (I^f_3-Q_f)\, s_W N_{i1}-I^f_3\, c_W
N_{i2}$, and the kinematic functions are given by
\begin{eqnarray}
&& D=(m_i+m_j)^2/((m_i+m_j)^2-m_Z^2)\,,\nonumber \\
&& F_{L,R}=(m_i+m_j)^2/(m^2_{\tilde{e}_{L,R}}+m_i m_j)\,,\nonumber\\
&& D^\prime=(m_i-m_j)^2/((m_i-m_j)^2 - m^2_Z)\,,\nonumber \\
&& F^\prime_{L,R}=(m_i-m_j)^2/(m^2_{\tilde{f}_{L,R}}-m_i m_j)\,.
\end{eqnarray}
In CP--invariant theories, all the (complex) rotation matrices
$R_{jk}$ in Eq.~(\ref{eq:Mdef}) become real and orthogonal.
Therefore, if the neutralinos $\tilde{\chi}^0_i$ and
$\tilde{\chi}^0_j$ have the same CP parity, then the Majorana
phase difference, $\alpha_{i}-\alpha_j$, is $0$ or $\pi$, and so
$N_{ik}N^*_{jl}$ is real. On the contrary, if the neutralino pair
have the opposite CP parity, the phase difference
$\alpha_i-\alpha_j$ is $\pm \pi/2$ and so $N_{ik}N^*_{jl}$ is
purely imaginary. Consequently, in CP--invariant theories the
cross section of a non--diagonal neutralino pair rises steeply in
S--waves only when the produced neutralinos have the opposite
parity, as dictated by the first CP relation (\ref{cpparity1}) and
as clearly indicated by Eq.~(\ref{eq:thres1}). One important
implication of the selection rule is that, even if the $\{ij\}$
and $\{ik\}$ pairs are excited steeply in S--waves, the pair
$\{jk\}$ must be excited slowly in P--waves characterized by the
slow rise $\sim\beta^3$ of the cross section \cite{petcov,ckmz}.
In contrast to the production case, the characteristic sharp
S--wave decrease of the fermion invariant mass distribution near
the end point is possible only if the neutralinos have the same CP
parity, as dictated by the second CP relation (\ref{cpparity2})
and as clearly indicated by Eq.~(\ref{eq:thres2}).\s

However, in the CP--noninvariant theories the orbital angular
momentum is no longer restricted by the selection rules
(\ref{cpparity1}) and (\ref{cpparity2}). The production of all
non--diagonal pairs can simultaneously be excited steeply in
S--waves near threshold, and the corresponding neutralino to
neutralino transition can be excited steeply in S--waves even if
the production cross section of the same non--diagonal neutralino
pair is excited steeply in S--waves. Consequently, CP violation in
the neutralino system can clearly be signalled by (i) the sharp
S--wave excitations of the production of three non--diagonal
$\{ij\}$, $\{ik\}$ and $\{jk\}$ pairs near threshold \cite{ckmz}
or by (ii) the simultaneous S--wave excitations of the production
of any non--diagonal $\{ij\}$ neutralino pair in $e^+e^-$
annihilation, $e^+e^-\rightarrow
\tilde{\chi}^0_i\tilde{\chi}^0_j$, near threshold and of the
fermion invariant mass distribution of the neutralino 3--body
decays, $\tilde{\chi}^0_i\rightarrow\tilde{\chi}^0_j\, f \bar{f}$,
near the end point.\s

It is noteworthy that only the light neutralinos
$\tilde{\chi}^0_{1,2}$ among the four neutralino states, which are
expected to be lighter than sfermions and gluino in many
scenarios, may be kinematically accessible in the initial phase of
$e^+e^-$ linear colliders. In this situation, the method based on
the threshold behaviors of the production of three different
non--diagonal neutralino pairs for probing CP violation is not
available. On the contrary, the combined analysis of the threshold
excitation of the production process, $e^+e^-\rightarrow
\tilde{\chi}^0_1 \tilde{\chi}^0_2$, and the fermion invariant mass
distribution of the decay,
$\tilde{\chi}^0_2\rightarrow\tilde{\chi}^0_1\, f\bar{f}$, near the
end point can still serve as one of the most powerful probes of CP
violation in the neutralino system even in the initial phase of
$e^+e^-$ linear colliders.\s

\begin{figure}[htb]
\begin{center}
\epsfig{file=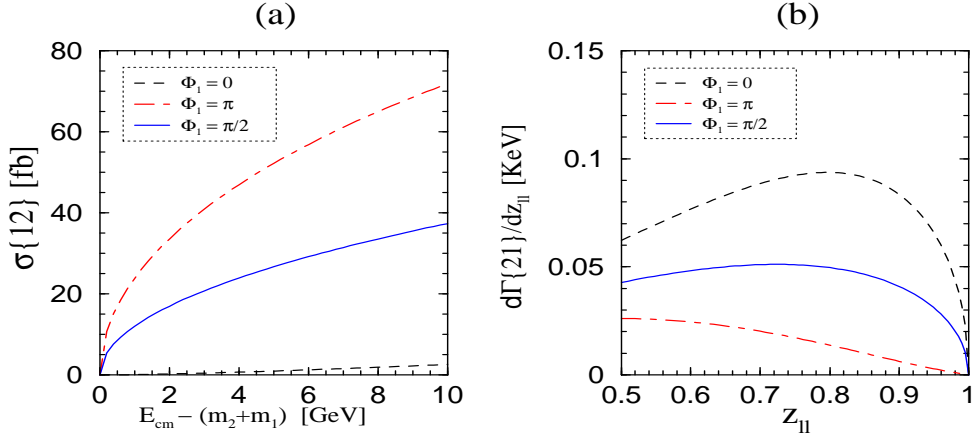,height=6cm,width=13cm} \caption{\it (a) The
         threshold behavior of the neutralino production
         cross--sections $\sigma{\{12\}}$ near the threshold and (b)
         the lepton invariant mass distribution of the decay
         $\tilde{\chi}^0_2\rightarrow\tilde{\chi}^0_1\, l^+l^-$ near the
         end point, illustrated for
         the parameter set: $\tan\beta=10$, $|M_1|=100$ GeV, $M_2=150$ GeV,
         $|\mu|=400$ GeV and $\Phi_\mu=0$ as well as the slepton masses,
         $m_{\tilde{l}_L}=250$ GeV and $m_{\tilde{l}_R}=200$ GeV.}
\label{fig:threshold}
\end{center}
\vskip -0.3cm
\end{figure}

In order to illustrate the method for probing CP violation
numerically, we take a parameter set for the fundamental SUSY
parameters\footnote{Analyses of electric dipole moments strongly
suggest that CP violation in the higgsino sector will be very
small in the MSSM if this sector is non--invariant at all
\cite{susycp,overcome}.}:
\begin{eqnarray}
\tan\beta =10;\quad |M_1|=100\, {\rm GeV},\quad M_2 = 150\, {\rm
GeV},\quad |\mu|=400\, {\rm GeV};\quad \Phi_\mu=0
\end{eqnarray}
and we choose two different values, $\{0, \pi\}$ for the phase
$\Phi_1$, in the CP--invariant case and one value, $\pi/2$, in the
CP non--invariant case. [The parameter point with such a large
phase $\Phi_1=\pi/2$ might already have been excluded by the
stringent EDM constraints. Nevertheless, this point is taken just
for illustrative purpose in the present work; the indirect EDM
limits depend also on many parameters of the theory outside the
neutralino sector.] We take the slepton masses,
$m_{\tilde{l}_L}=250$ GeV and $m_{\tilde{l}_R}=200$ GeV and
consider the 3--body leptonic decay
$\tilde{\chi}^0_2\rightarrow\tilde{\chi}^0_1\, l^+l^-$, especially
with $l=e, \mu$, for the illustration. We note that the
neutralinos $\tilde{\chi}^0_1$ and $\tilde{\chi}^0_2$ have the
same (opposite) CP parity for $\Phi_1=0$ ($\Phi_1=\pi$). As
expected from the selection rules (\ref{cpparity1}) and
(\ref{cpparity2}) in the CP--invariant case,
Figure~\ref{fig:threshold} clearly shows that if the production of
the neutralino pair $\tilde{\chi}^0_1\tilde{\chi}^0_2$ in $e^+e^-$
annihilation increases slowly in P--waves (steeply in S--waves)
near threshold, then the lepton invariant mass distribution of the
decay $\tilde{\chi}^0_2\rightarrow\tilde{\chi}^0_1\, l^+l^-$
decreases steeply in S--waves (slowly in P--waves) near the end
point for the neutralino pair of the same (opposite) CP parity
with $\Phi_1=0$ ($\Phi_1=\pi$). On the contrary, in the
CP--noninvariant case ($\Phi_1=\pi/2$) the production and decay
are excited steeply both in S--waves.

\section{Conclusions}
\label{sec:conclusion}

We have shown that only in CP--noninvariant theories the
production of any non--diagonal neutralino pair
$\tilde{\chi}^0_i\tilde{\chi}^0_j$ ($i\neq j$) in $e^+e^-$
annihilation near threshold and the fermion invariant mass
distribution of the 3--body neutralino fermionic decay
$\tilde{\chi}^0_i\rightarrow \tilde{\chi}^0_j\, f\bar{f}$ near the
end point can simultaneously be excited steeply in S--waves.\s

In light of the possibility that only the two light neutralinos
$\tilde{\chi}^0_1$ and $\tilde{\chi}^0_2$ among the four
neutralinos can be accessed kinematically in the initial phase of
$e^+e^-$ linear colliders, the combined analysis of the production
of the neutralino pair $\tilde{\chi}^0_1\tilde{\chi}^0_2$ in
$e^+e^-$ annihilation near threshold and the neutralino decay
$\tilde{\chi}^0_2\rightarrow \tilde{\chi}^0_1\, f\bar{f}$ near the
end point of its fermion invariant mass could provide a first
qualitative indication of the CP violation in the neutralino
system. \vskip 0.5cm

\subsection*{Acknowledgments}
The author would like to thank M. Drees, J. Kalinowski and P.M.
Zerwas for helpful comments. The work was supported in part by the
Korea Research Foundation Grant (KRF--2002--070--C00022) and in
part by KOSEF through CHEP at Kyungpook National University.

\end{document}